\documentclass{aa}  
\usepackage{graphicx}
\usepackage{txfonts}
\usepackage{color}

\newcommand\gaia{\textit{Gaia}\xspace}
\newcommand\gdrone{\gaia~DR1\xspace}
\newcommand\gdrtwo{\gaia~DR2\xspace}

\newcommand{\VT}{$V_\mathrm{T}$\xspace}
\newcommand{\msol}{M$_\odot$\xspace}

\newcommand{\gbp}{{G$_\mathrm{BP}$}\xspace}
\newcommand{\grp}{{G$_\mathrm{RP}$}\xspace}

\newcommand{\mg}{M$_{\mathrm{G}}$\xspace}
\newcommand{\mj}{M$_{\mathrm{J}}$\xspace}
\newcommand{\mh}{M$_{\mathrm{H}}$\xspace}
\newcommand{\mk}{M$_{\mathrm{Ks}}$\xspace}

\usepackage{ulem}
\usepackage{amstext}

\begin{document}

   \title{New ultra-cool and brown dwarf candidates in \gdrtwo}

   \subtitle{}

   \author{C. Reyl\'e
          \inst{1}
          }

   \institute{Institut UTINAM, CNRS UMR6213, Univ. Bourgogne Franche-Comt\'e, OSU THETA Franche-Comt\'e-Bourgogne, Observatoire de Besan\c con, BP 1615, 25010 Besan\c con Cedex, France.\\
              \email{celine@obs-besancon.fr}
             }

   \date{Received ; accepted }

 
  \abstract
   {The second Gaia data release (\gdrtwo) contains high-precision positions, parallaxes, and proper motions for 1.3 billion
sources. The resulting Hertzsprung-Russel diagram reveals fine structures throughout the mass range.
   }
 {This paper aims to investigate the content of \gdrtwo\ at the low-mass end and to characterize ultra-cool and brown dwarfs. 
 }
  {We first retrieved the sample of spectroscopically confirmed ultra-cool and brown dwarfs in \gdrtwo. We used their locus in the precise Hertzsprung-Russel diagram to select new candidates and to investigate their properties. 
  }
   {The number of spectroscopically confirmed objects recovered in \gdrtwo corresponds to 61\% and 74\% of the expected number of objects with an estimated \gaia magnitude G$_\mathrm{est}\leq21.5$ and 20.3, respectively. This fills much of the gap to \gdrone. Furthermore, \gdrtwo\ contains $\sim$13 000 $\geq$M7 and 631 new L candidates. A tentative classification suggests that a few hundred of them are young or subdwarf candidates. Their distance distribution shows that the solar neighborhood census is still incomplete.
   }
   {\gdrtwo\ offers a great wealth of information on low-mass objects. It provides a homogeneous and precise catalog of candidates that is worthwhile to be further characterized with spectroscopic observations.
   }


\maketitle

\section{Introduction}

The \gaia\ mission \citep{Prusti2016} provides  full-sky coverage down to the \gaia\ magnitude G = 20 (V$\simeq$20-22) at the spatial resolution of the HST, with about 70 observations per source over five years. Distances, with an amazing precision of 1\% up to 2.5 kpc at the end of the mission, and proper motions are now  measured for 1.3 billions stars.

As highlighted in the scientific demonstration paper from \cite{Babusiaux2018}, \gaia\ astrometry and photometry are powerful tools for studying fine structures within the Hertzsprung-Russell (HR) diagram. It is now possible to distinguish the locus of different types of objects, as predicted by models.

The \gaia\ optical observations are not the most suitable observations for studying the low-mass reddest end of the HR diagram. Nevertheless, the large number of objects observed with an unprecedented precision, including trigonometric parallax, makes it an unrivaled dataset for studying the ultra-cool and brown dwarfs.

These low-mass objects are the dominant stellar component of the Milky Way \citep[$\sim$ 70\% of all stars, ][]{Reid1997,Bochanski2010}. They still remain elusive, however, because of their low luminosity, and modeling their complex atmosphere is still a challenge.
Therefore, the understanding of this population has
relevant implications for both stellar and Galactic astronomy.

Ultra-cool dwarfs have been defined by \cite{Kirkpatrick1997} as M7 and later main-sequence stars. This corresponds to an effective temperature of $\lesssim$ 2700 K \citep{Rajpurohit2013}. These objects serve as a link between known stars and brown dwarfs. They span the transition from stellar to sub-stellar masses: this effective temperature is expected in a 0.03 \msol\ object of 8 Myr or in a 0.095 \msol\ object of 10 Gyr \citep{Baraffe2015}, both at solar metallicity.   

\cite{Smart2017} compiled an input catalog of known L and T dwarfs with an estimated G magnitude within the reach of \gaia. They 
identified part of them in \gdrone\ \citep{Brown2016}. They found 321 L and T dwarfs, with 10 $\geq$ L7, which corresponds to 45\% of the brown dwarfs with estimated \gaia magnitude G$_\mathrm{est}\leq20.3$. As they pointed out, this incompleteness is mainly due to the cuts made in the catalog to ensure the quality of the data.
With eight more months of observations, \gdrtwo\ \citep{Brown2018} should provide more complete and precise data, and help to fill this gap. Recent discoveries show that even the local census is not complete, as illustrated by the discovery of an L7 at 11 pc, close to the Galactic plane \citep{Scholz2018,Faherty2018}.

In a first step, we search for known ultra-cool and brown dwarfs in \gdrtwo\ and investigate their properties using their \gaia\ observations (section~\ref{known}). In a second step, we use the properties from the known sample to show the potential of \gdrtwo\ to reveal new candidates and list them (section~\ref{new}). The conclusions and perspectives are given in section~\ref{ccl}.

\section{Known ultra-cool and brown dwarfs found in \gdrtwo}
\label{known}

\subsection{Sample}

We focus on the well-characterized sample of ultra-cool and brown dwarfs that have a spectroscopic spectral type. A compilation of all L, T, and Y dwarfs that \gaia is expected to observe directly or constrain indirectly (e.g., in a common proper motion system with brighter members), the so-called \gaia Ultracool Dwarf Sample (GUCDS), has been published by \cite{Smart2017}. GUCDS is based on the online census compiled by J. Gagn\'e\footnote{https://jgagneastro.wordpress.com/list-of-ultracool-dwarfs/} , which contains all objects from the Dwarfarchives\footnote{http://dwarfarchives.org} database, complemented by the catalogs of \cite{Dupuy2012} and \cite{Mace2014}. Some of these objects have information on age or metallicity, and they are listed in the census as young candidates or subdwarfs. \cite{Smart2017} added the recent discoveries from \cite{Marocco2015} and \cite{Faherty2016}. This catalog has 1886 entries, for which the predicted G magnitude is brighter than 21.5 for 1010 L and 58 T dwarfs. This is within the reach of \gaia.

To this census, we added the ultra-cool (M7-M9.5) dwarfs listed by J. Gagn\'e\footnote{https://jgagneastro.wordpress.com/list-of-m6-m9-dwarfs/} and the new spectroscopically confirmed objects from \textcolor[rgb]{0.992157,0.501961,0.0313726}{\cite{Rajpurohit2014,Robert2016,Faherty2018,Zhang2018}.}  Because we wish to study the location in the fine \gdrtwo HR diagram of a well-characterized sample, we did not include the numerous candidates that are found in photometric surveys without spectroscopic confirmation.

We retrieved these entries using their 2MASS identifier and cross-matching \gdrtwo\ with the 2MASS catalog \citep{Skrutskie2006} provided in the \gaia\ archive
\citep{Marrese2017}. We also used the \gdrtwo\ identifier when available in the SIMBAD astronomical database\footnote{http://simbad.u-strasbg.fr/simbad/} \citep{Wenger2000}, particularly for the faint objects found within the SDSS \citep{Eisenstein2011} and SIMPS \citep{Artigau2009} catalogs that have no counterpart in 2MASS. 

\begin{figure}[ht!]
\begin{center}
\includegraphics[height=7.5cm]{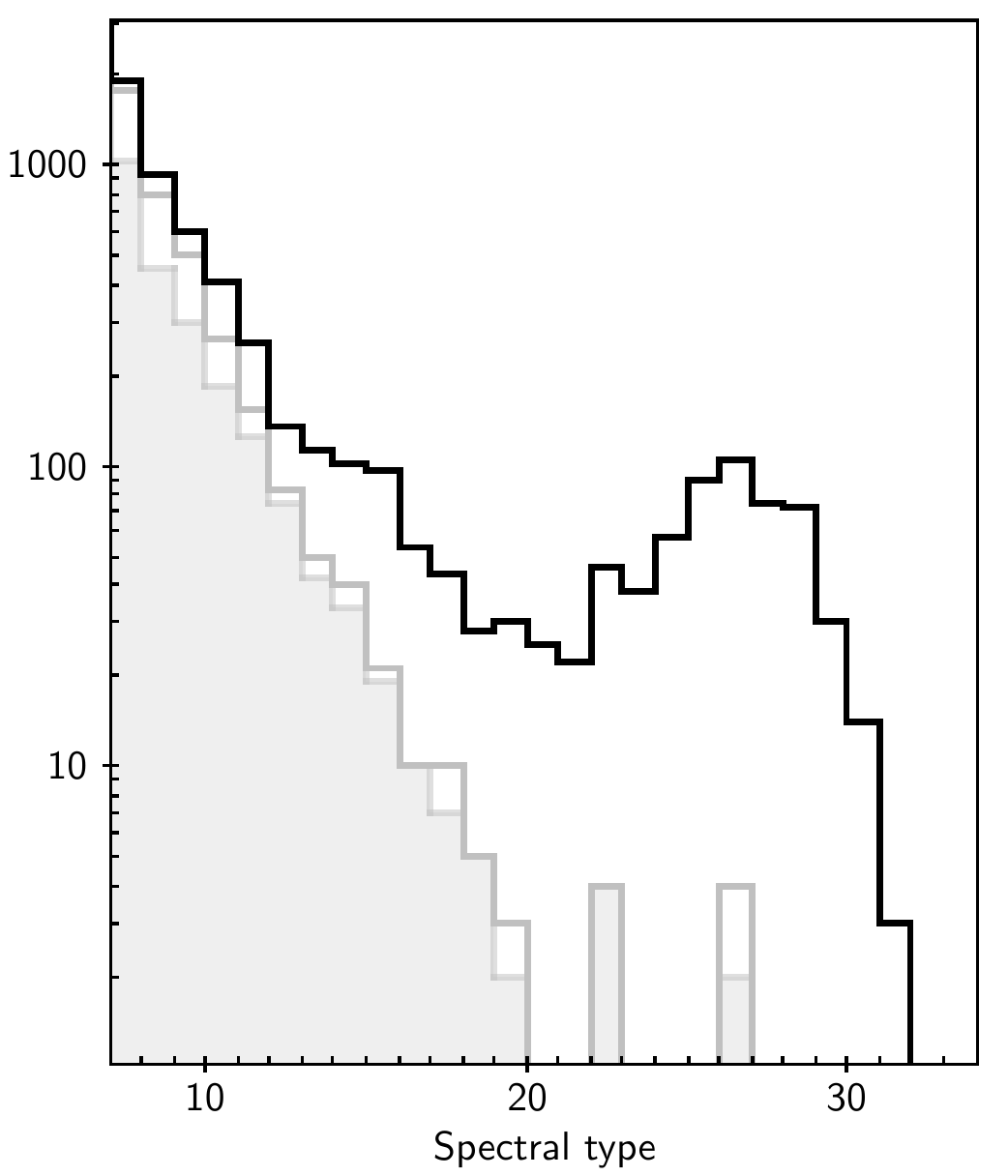} 
\caption{Spectral type distribution of the known ultra-cool ($\geq$M7) and brown dwarfs. Black line: Input catalog. Gray line: Objects with a counterpart in \gdrtwo. Filled gray: Objects with a counterpart in \gdrtwo and $\sigma_\varpi \leq$ 10\%. 10 stands for L0, 20 for T0, and 30 for Y0 on the spectral type axis.}
\label{f:histmltspt}
\end{center}
\end{figure}

The resulting sample includes 3671 ultra-cool ($\geq$M7), 647 L, and 16 T dwarfs, 34 of which have a spectral type $\geq$ L7. Of these, 1806 $\geq$M7, 526 L, and 9 T dwarfs have a relative precision on parallax $\sigma_\varpi \leq$ 10\%, and full astrometric (coordinates, parallax, and proper motions) and photometric (G, \gbp, and \grp) information. Twenty-six are noted in the census as subdwarfs and 75 as young candidates. The spectral type distribution of the input catalog, the sample with an entry in \gdrtwo, and the sub-sample with $\sigma_\varpi \leq$ 10\%  is shown in Fig.~\ref{f:histmltspt}. 

We built the HR diagram by computing the absolute \gaia magnitude in the G band for individual stars using $M_G = G-5 \log_{10}(1000/\varpi)+5$, where $\varpi$ is the parallax in milli-arcseconds. The simple distance determination from $\varpi$  is valid when $\sigma_\varpi \lesssim$ 20\% \citep{Luri2018}.

\begin{figure}[ht!]
\begin{center}
\includegraphics[height=7.5cm,clip=]{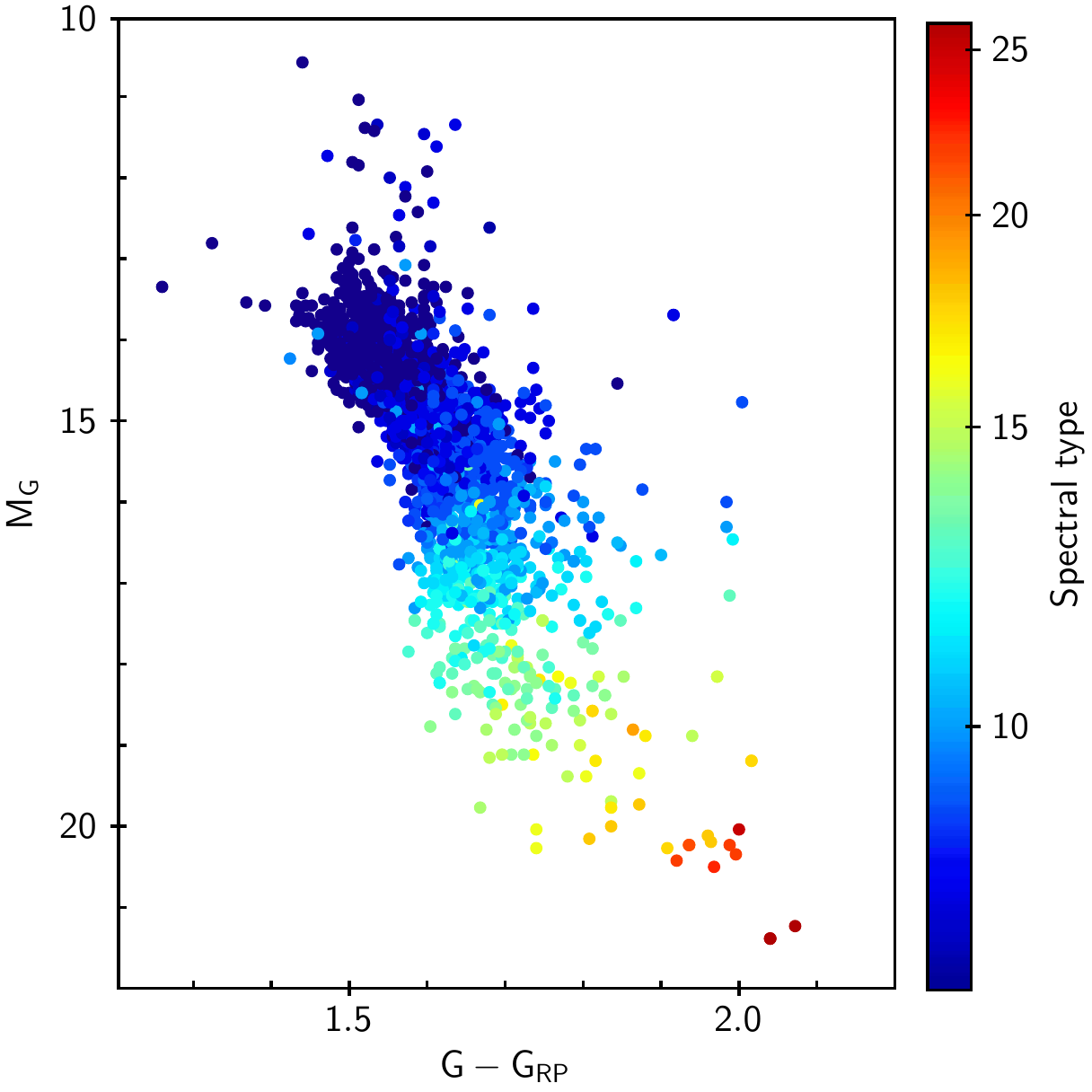} 
\caption{HR diagram of the known ultra-cool and brown dwarfs found in \gdrtwo\ that have $\sigma_\varpi \leq$ 10\%. 
The color bar gives the spectral type range, where 10 stands for L0 and 20 for T0.}
\label{f:hrdmltspt}
\end{center}
\end{figure}

Fig.~\ref{f:hrdmltspt} shows the HR diagram of the sub-sample with $\sigma_\varpi \leq$ 10\%. The color code gives the spectral type. The adopted spectral type is the mean of the optical and near-infrared spectral type when both are available. We used G-\grp\ as the color index because these faint and red objects have a lower flux in the BP bandpass. 

\begin{figure*}[ht!]
\begin{center}
\includegraphics[height=7.5cm,clip=, viewport=0 0 288 350]{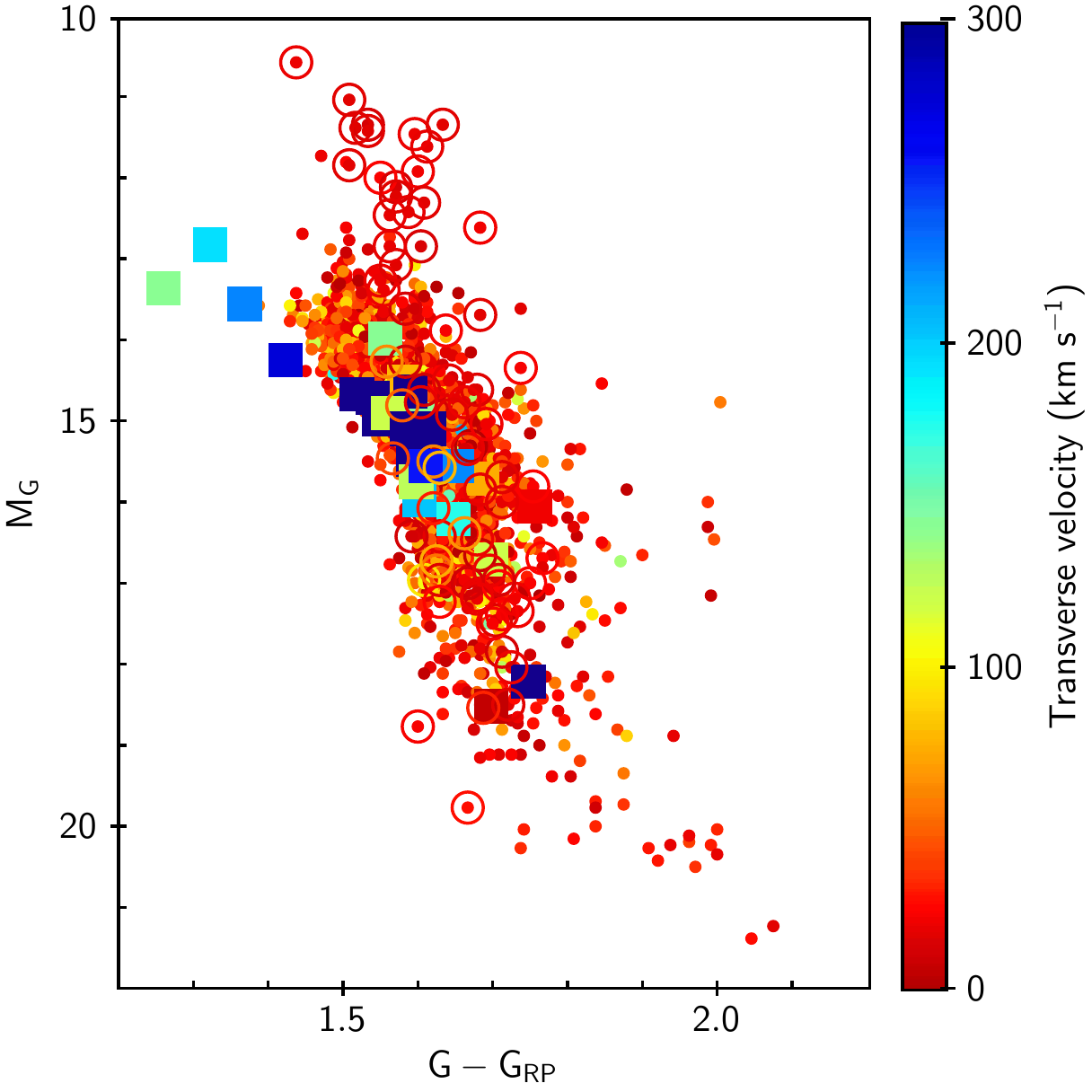}   
\includegraphics[height=7.5cm]{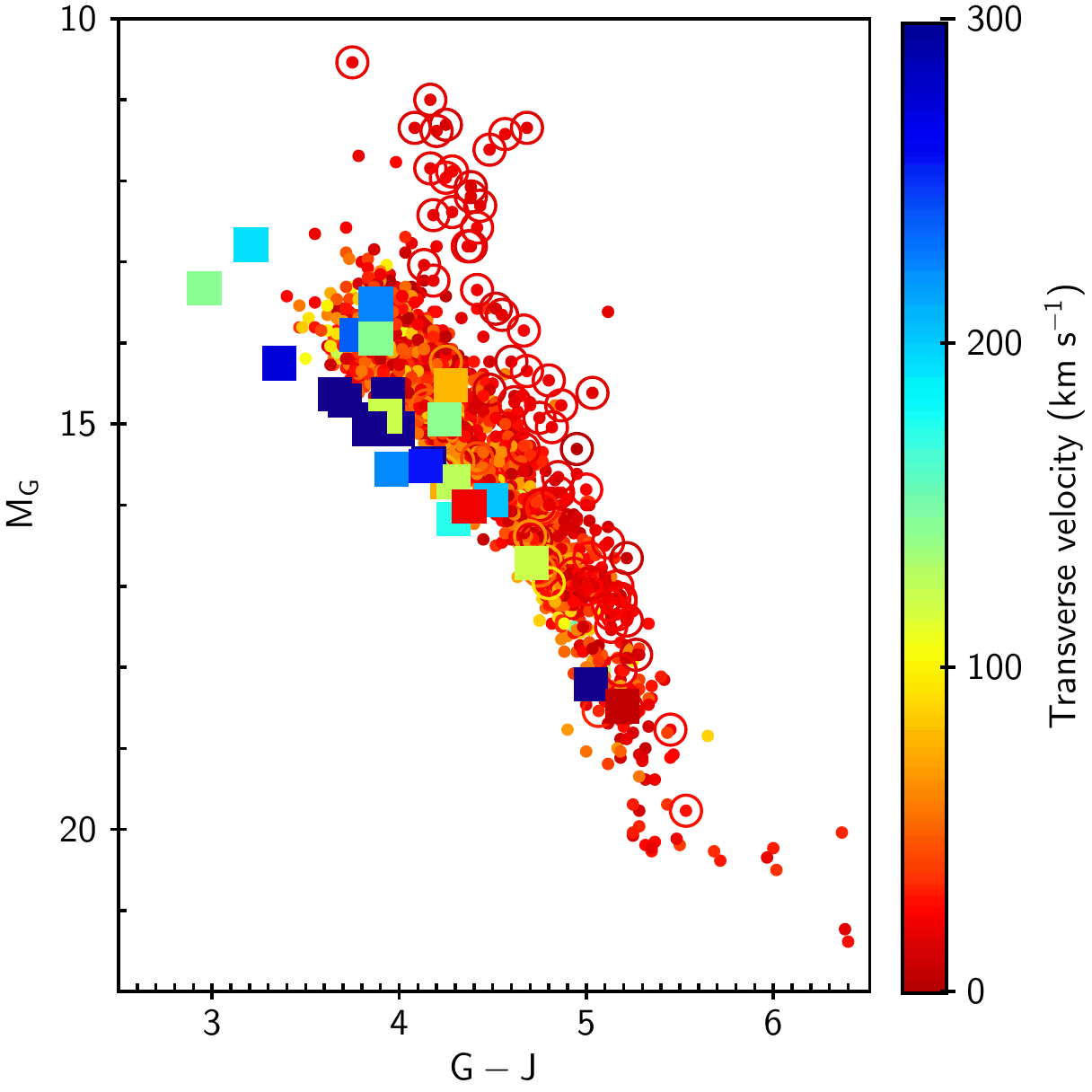}   
\caption{HR diagram of the known sample found in \gdrtwo (restricted to the sub-sample with $\sigma_\varpi \leq$ 10\%), using the G-\grp (left) and G-J (right) color index. The color bar gives the transverse velocity range. Filled squares are objects listed as subdwarfs in the census, and open circles are objects listed as young candidates.}
\label{f:hrdmltvtan}
\end{center}
\end{figure*}

As the sub-sample with $\sigma_\varpi \leq$ 10\% contains all astrometric properties, it allows us to explore the HR diagram as a function of transverse velocity. The transverse velocity \VT was computed as follows: \VT $= \sqrt{\mu_\alpha^2+\mu_\delta^2} \times \frac{4.74}{\varpi}$, where $\mu_\alpha$ and $\mu_\delta$ are the proper motions. This is shown in Fig.~\ref{f:hrdmltvtan}.
Thanks to the high precision of \gaia\ measurements, it is possible to distinguish a sequence of young candidates (redder and brighter) with a low transverse velocity from the sequence of lower metallicity dwarfs (bluer and fainter). The vast majority of the subdwarfs have a high transverse velocity, indicating that they belong to the older populations of the Milky Way, as expected. This separation seems slightly enhanced in the G-J color index over the G-\grp\ index.

\subsection{Color, absolute magnitude, and spectral type relations}

\begin{figure*}[ht!]
\begin{center}
\includegraphics[height=7.5cm]{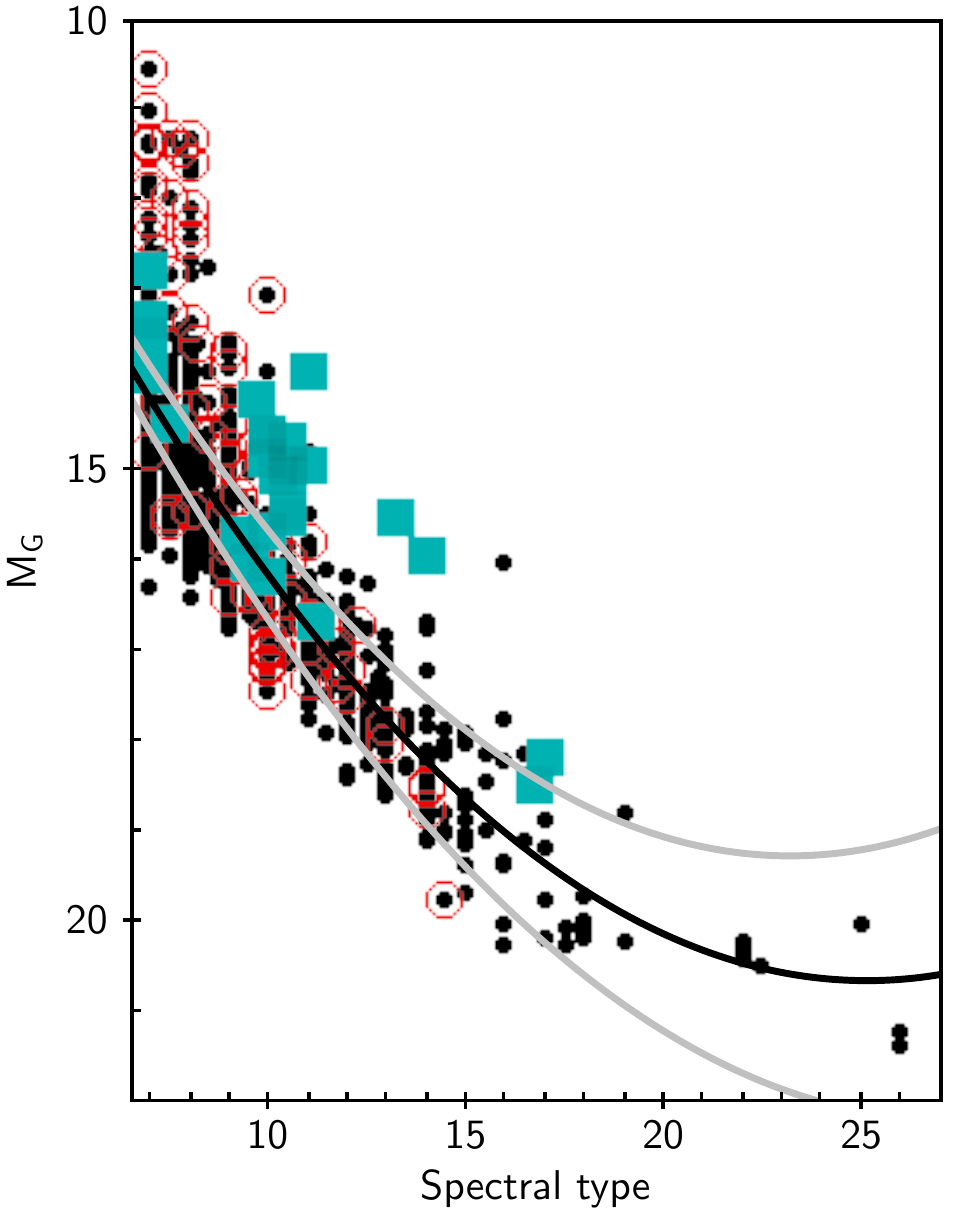} 
\includegraphics[height=7.5cm]{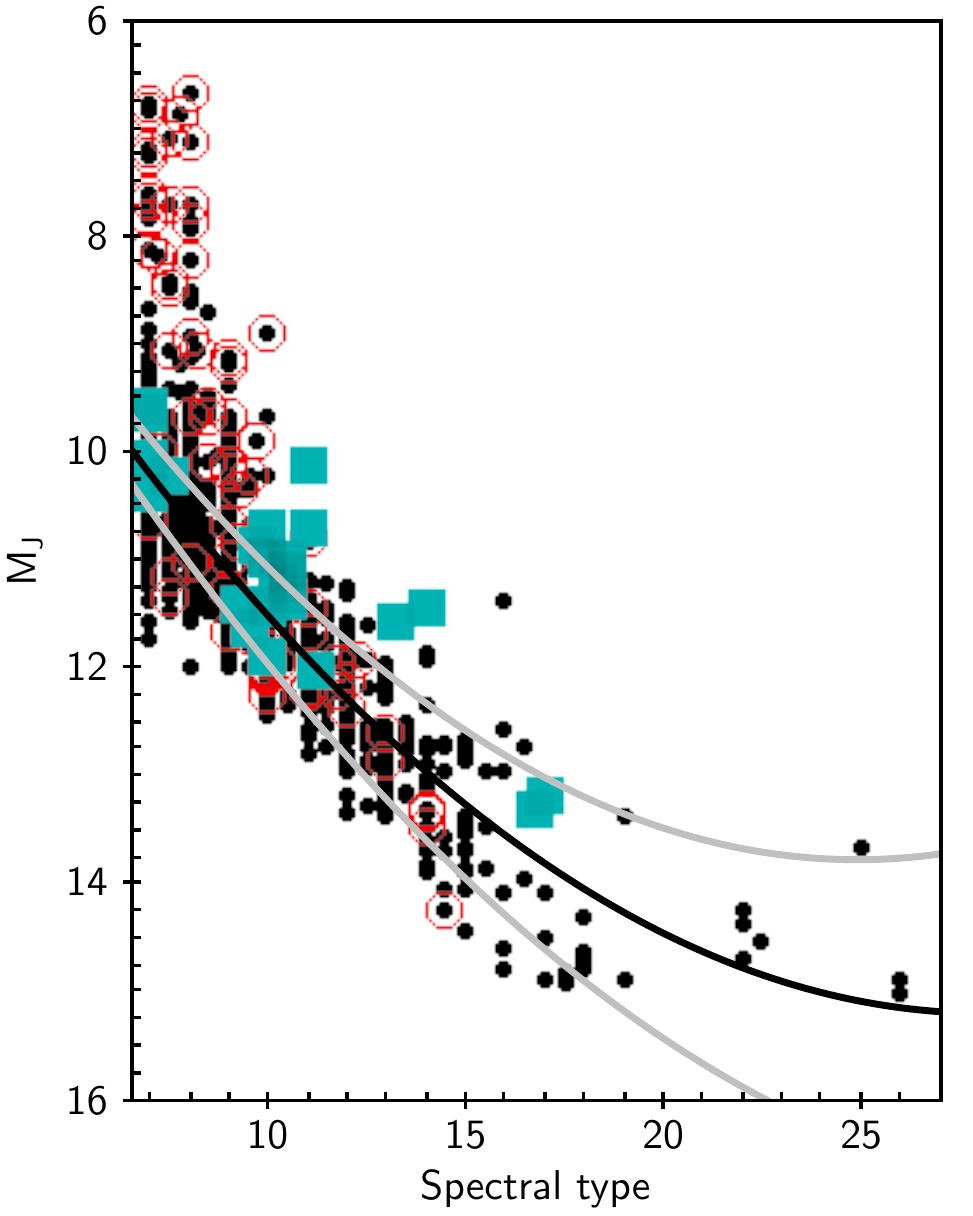} 
\includegraphics[height=7.5cm]{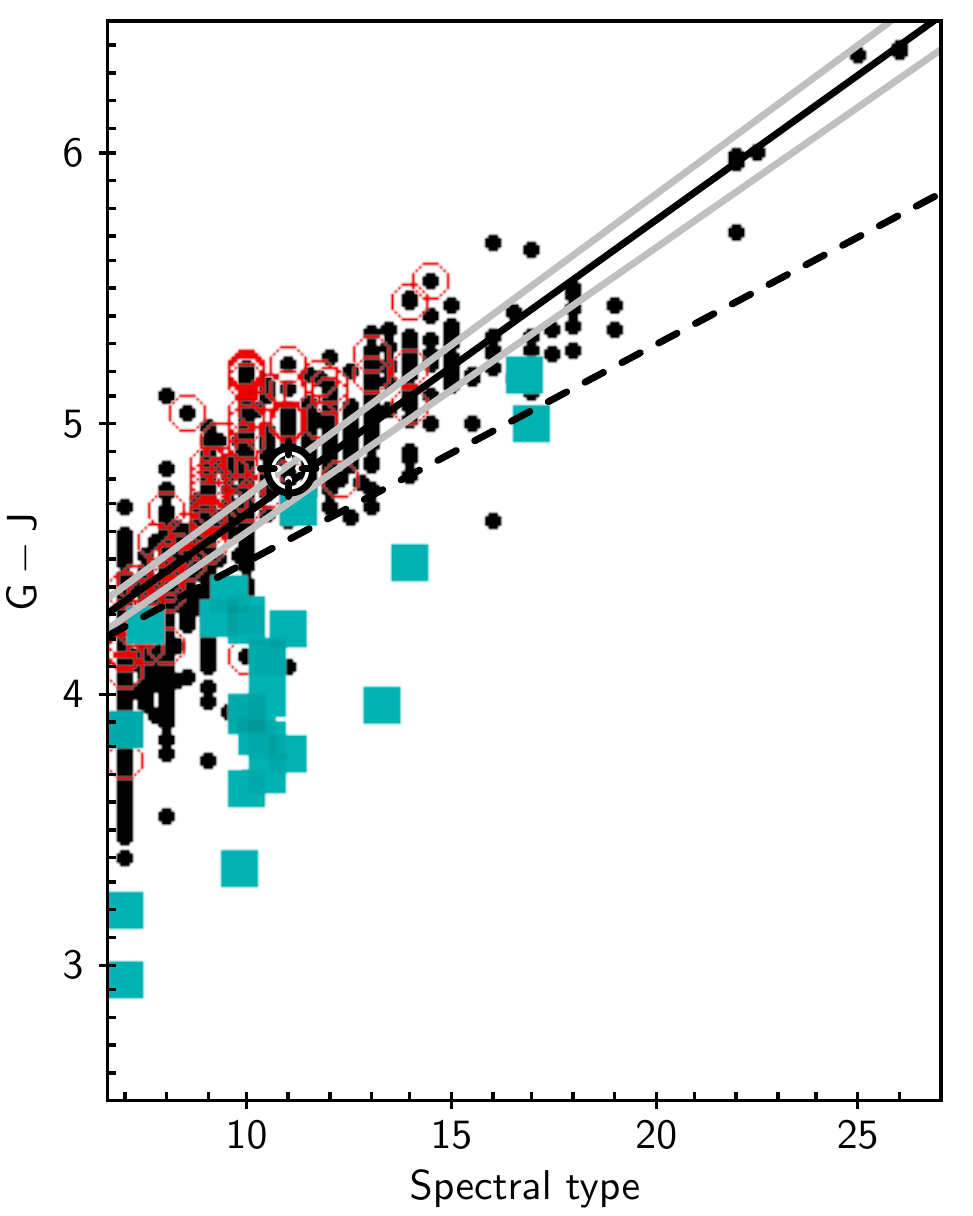} \caption{M$_{\mbox{G}}$ (left), M$_{\mbox{J}}$ (middle)}, and G-J (right) relations vs. spectral type of the known ultra-cool and brown dwarfs found in \gdrtwo. 10 stands for L0, and 20 for T0. The squares show the subdwarfs, and the open circles the young candidates. The black line gives the  fits, and the gray lines include the  asymptotic standard errors. The G-J vs. spectral type relation is valid from M7 to L9). The black dashed line is the relation derived from L0 to L7 found in \gdrone \citep{Smart2017}.
\label{f:mgspt}
\end{center}
\end{figure*}

We determined the relationships between \mg, \mj, \mh, \mk, G-J, J-K$_{\mbox{s}}$ , and spectral type from a polynomial fitting of the sub-sample ( $\sigma_\varpi \leq$ 10\%). We also determined the \mj versus J-K$_{\mbox{s}}$ relation. The resulting parameters and their asymptotic standard errors are summarized in Table~\ref{t:relations}. The validity range of each relation is also given. 
Some of the relations are shown in Fig.~\ref{f:mgspt}. 
\begin{table*}[ht]
\caption{Colors, absolute magnitudes, and spectral type relations obtained from the polynomial fitting Y = aX$^2$+bX+c. \mbox{SpT} ranges from 7 (for M7) to 26 (for T6). We list the range in which the relations are valid.}
\begin{tabular}{llllll}
\hline\\
 Y & X &X-range &a &b &c\\
\hline\\
\mg & SpT &M7-T6 &-0.020 $\pm$ 0.001 &0.985 $\pm$ 0.025 &8.298 $\pm$ 0.135\\
\mj & SpT &M7-T6 &-0.011  $\pm$ 0.001 &0.627 $\pm$ 0.023 &6.372 $\pm$ 0.122 \\
\mh & SpT &M7-T6 &-0.007 $\pm$ 0.001 &0.503 $\pm$ 0.023 &6.453 $\pm$ 0.122 \\
\mk & SpT &M7-T6 &-0.003 $\pm$ 0.001 &0.387 $\pm$ 0.023 &6.709 $\pm$ 0.122\\
G-\grp & SpT &M7-T6 &0 &0.029 $\pm$ 0.001 &1.358 $\pm$ 0.005\\
G-J & SpT  &M9-T6 &0 &0.108 $\pm$ 0.003 &3.603 $\pm$ 0.036\\
J-K$_{\mbox{s}}$ & SpT &M7-L9 &0 &0.0635 $\pm$ 0.002  &0.562 $\pm$ 0.016\\
\mj & J-K$_{\mbox{s}}$ &0.5-2.2 &2.119 $\pm$ 0.116 &-2.531 $\pm$ 0.282 &10.967 $\pm$ 0.172\\
\hline\\
\end{tabular}
\label{t:relations}
\end{table*}

The dashed line in the right panel of Fig.~\ref{f:mgspt} represents the relation derived from 304 L dwarfs found in \gdrone \citep{Smart2017}. The difference is due to the evolution of the photometric system between \gdrone and \gdrtwo, as described in \cite{Brown2018}.

Young candidates depart from these relationships, in particular in the M-dwarf regime. It is also possible to distinguish the relation followed by the subdwarfs. Such trends have been presented extensively in different photometric systems by \cite{Zhang2018} based on their sample of 20 L subdwarfs found in \gdrtwo.

\section{New ultra-cool and brown dwarf candidates in \gdrtwo}
\label{new}

Our aim is to search for robust candidates in \gdrtwo, in particular, to define a list of objects deserving further investigations such as spectroscopic follow-up. Thus we chose to apply strict filters on the data that were based on astrometric and photometric features. Detailed information on the data processing, validation, and catalog content are given by \cite{Lindegren2018} for the astrometry and by \textcolor[rgb]{0.992157,0.501961,0.0313726}{\cite{Riello2018,Evans2018}} for the photometry. The global validation of the catalog can be found in \cite{Arenou2018}, and the description of the \gaia archive\footnote{https://gea.esac.esa.int/archive/}  in \cite{Salgado2018}.

\subsection{Data filtering}
 
In order to work on a precise HR diagram, we followed \cite{Babusiaux2018} to filter the data.
We refer to this paper for a detailed description of the applied filters. Here we describe their effects only briefly. The Gaia collaboration retained objects with $\sigma_\mathrm{M_{G}}<0.22$ mag, $\sigma_\mathrm{G}<0.022$ mag, $\sigma_\mathrm{G_{RP}}<0.054$ mag, and reliable five-parameter solutions (astrometric and photometric), and removed most artifacts. 

The corresponding \gaia\ archive query  is
{\small
\begin{verbatim}
SELECT * FROM gaiadr2.gaia_source 
WHERE parallax_over_error > 10
AND phot_g_mean_flux_over_error>50
AND phot_rp_mean_flux_over_error>20
AND visibility_periods_used>8
AND astrometric_chi2_al/(astrometric_n_good_obs_al-5)
   <1.44*greatest(1,exp(-0.4*(phot_g_mean_mag-19.5)))
\end{verbatim}
}

Following \cite{Babusiaux2018}, we also restricted the sample to low-extinction regions in the Milky Way and kept objects with a reddening $E(B-V) \leq 0.015$ according to the 3D extinction map of \cite{Capitanio2017}. The resulting HR diagram is shown in the left panel of Fig.~\ref{f:hrddr2}. 

\subsection{Color excess}
The HR diagram has a fuzzy appearance between the white dwarfs and the main sequence, and redward of the main sequence. 
Whereas the G flux is determined from a profile fitting, the fluxes in the blue (BP) and red (RP) photometers
are the total flux in a field of 3.5 $\times$ 2,1 arcsec$^2$.  Therefore, the measured BP and RP fluxes may include a contribution of flux from bright sky background (e.g.,  in crowded regions or from a nearby
source). As explained  in \cite{Evans2018}, I$_\mathrm{G}$, I$_\mathrm{BP}$, and I$_\mathrm{RP}$, the flux in the G, \gbp, and \grp bandpass, respectively, should be consistent (the sum of the BP and RP fluxes is expected to exceed the G flux by only a small factor).
They defined the color excess factor \texttt{phot\_bp\_rp\_excess\_factor} $=(\mathrm{I}_\mathrm{BP}+\mathrm{I}_\mathrm{RP})/\mathrm{I}_\mathrm{G}$ and an empirical limit to reject the affected sources having \texttt{phot\_bp\_rp\_excess\_factor} $\geq 1.3 + 0.06 \times (\mathrm{G}_\mathrm{BP}-\mathrm{G}_\mathrm{RP})^2$.

\begin{figure*}[ht!]
\begin{center}
\includegraphics[height=7.5cm]{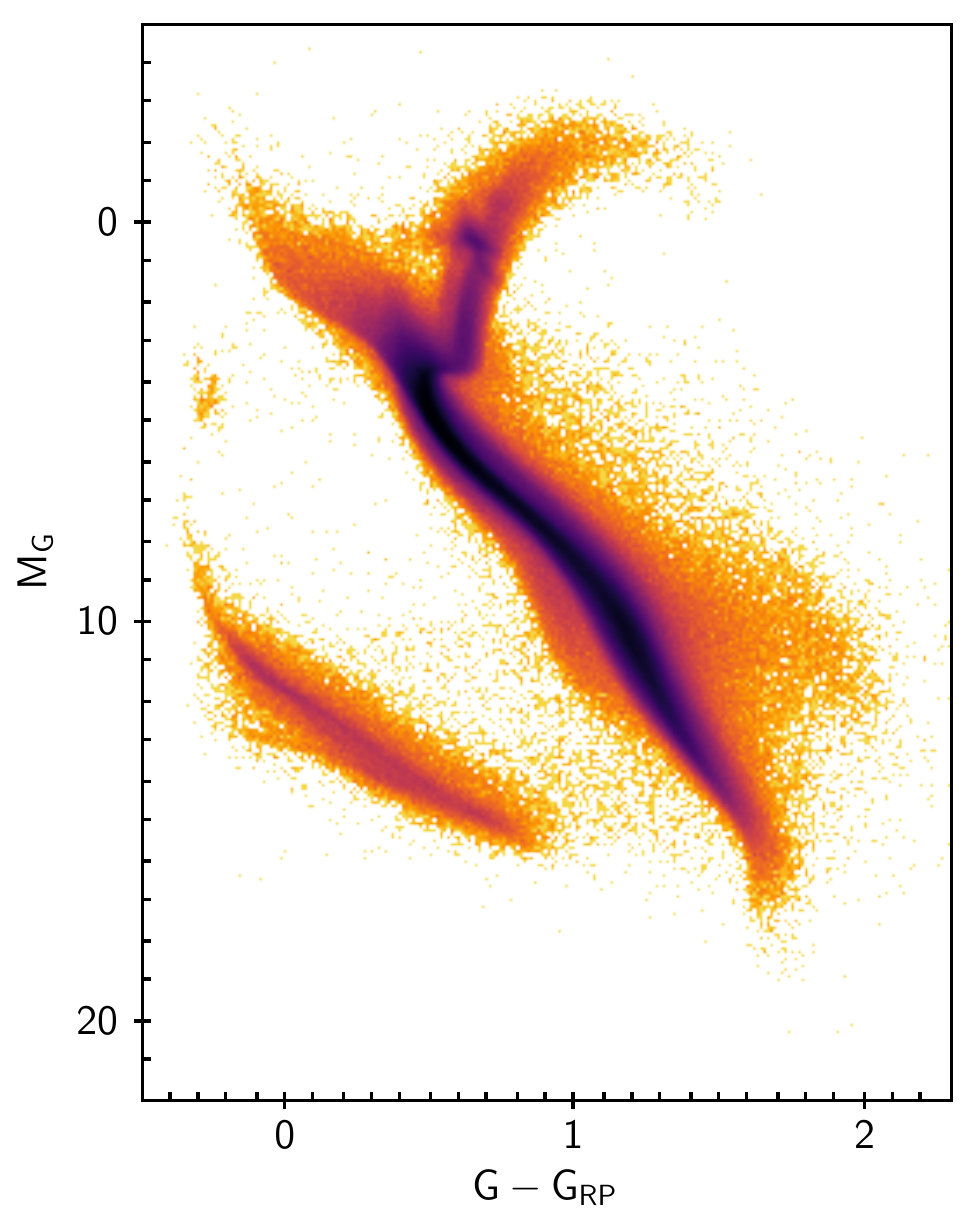}   
\includegraphics[height=7.5cm]{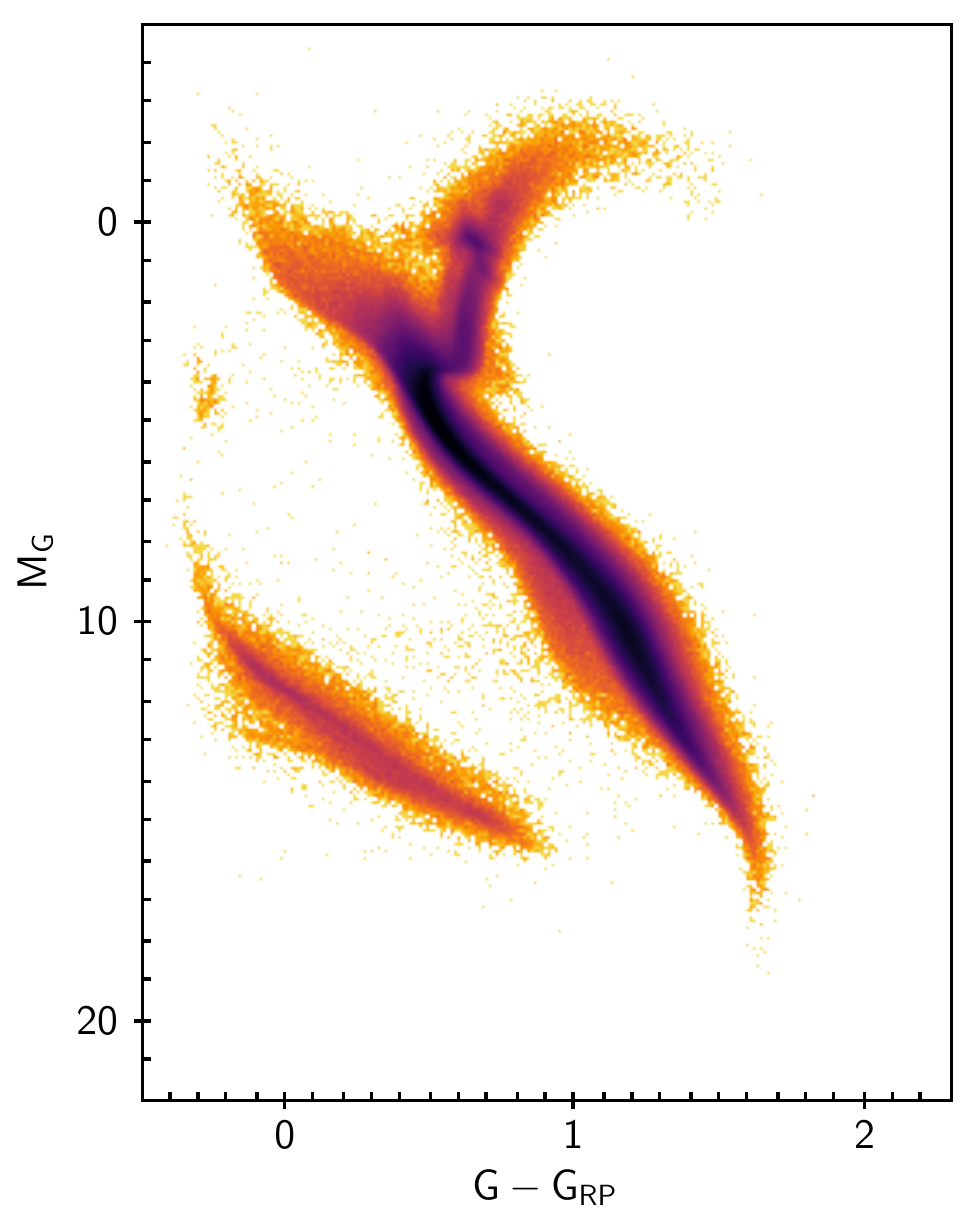}   
\includegraphics[height=7.5cm]{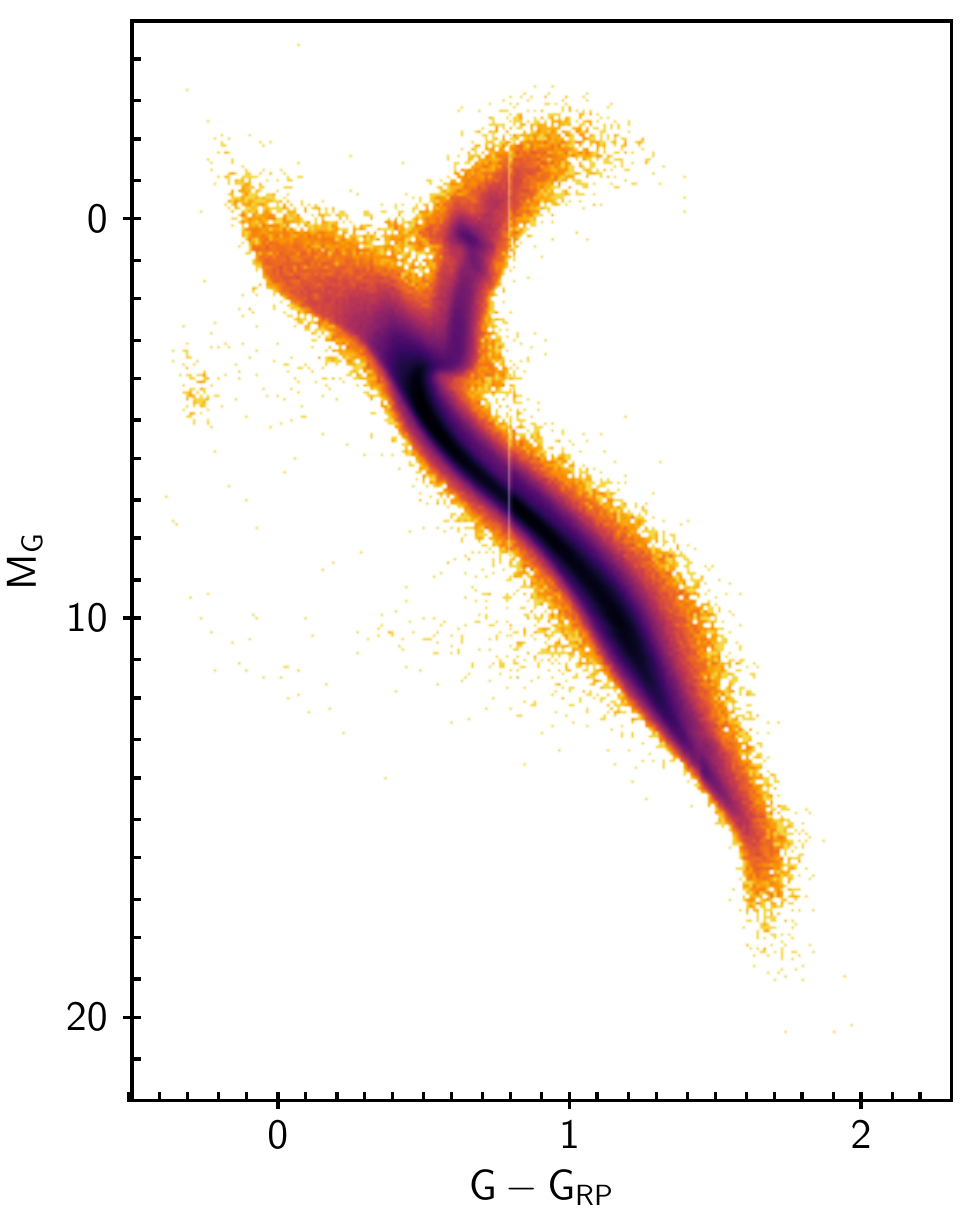}   
\caption{HR diagram of \gdrtwo. Left: 4 705 366 stars, after filtering, and in low-extinction Galactic regions. Middle: 4 640 635 stars, after removing stars with an RP/BP flux excess following \cite{Evans2018}. Right: 3 716 407 stars with the 2MASS photometric quality flag Q$_\mathrm{fl}$ = AAA, and after removing stars showing color excess using G-J (see text).}
\label{f:hrddr2}
\end{center}
\end{figure*}

The resulting HR diagram when applying this filter is shown in the middle panel of Fig.~\ref{f:hrddr2}. Because ultra-cool and brown dwarfs have low flux in the
BP bandpass, they are very sensitive to any background overestimation yielding a high \texttt{phot\_bp\_rp\_excess\_factor}. As a consequence, most of these objects are rejected, as can be clearly seen at the low-mass end of the main sequence. As an illustration, applying this filter on the previously known sample dramatically removes 47\% of the objects (68\% of the L and T dwarfs) that were found in \gdrtwo (1067 ultra-cool and only  160 L dwarfs remain).

\begin{figure*}[ht!]
\begin{center}
\includegraphics[height=7.5cm]{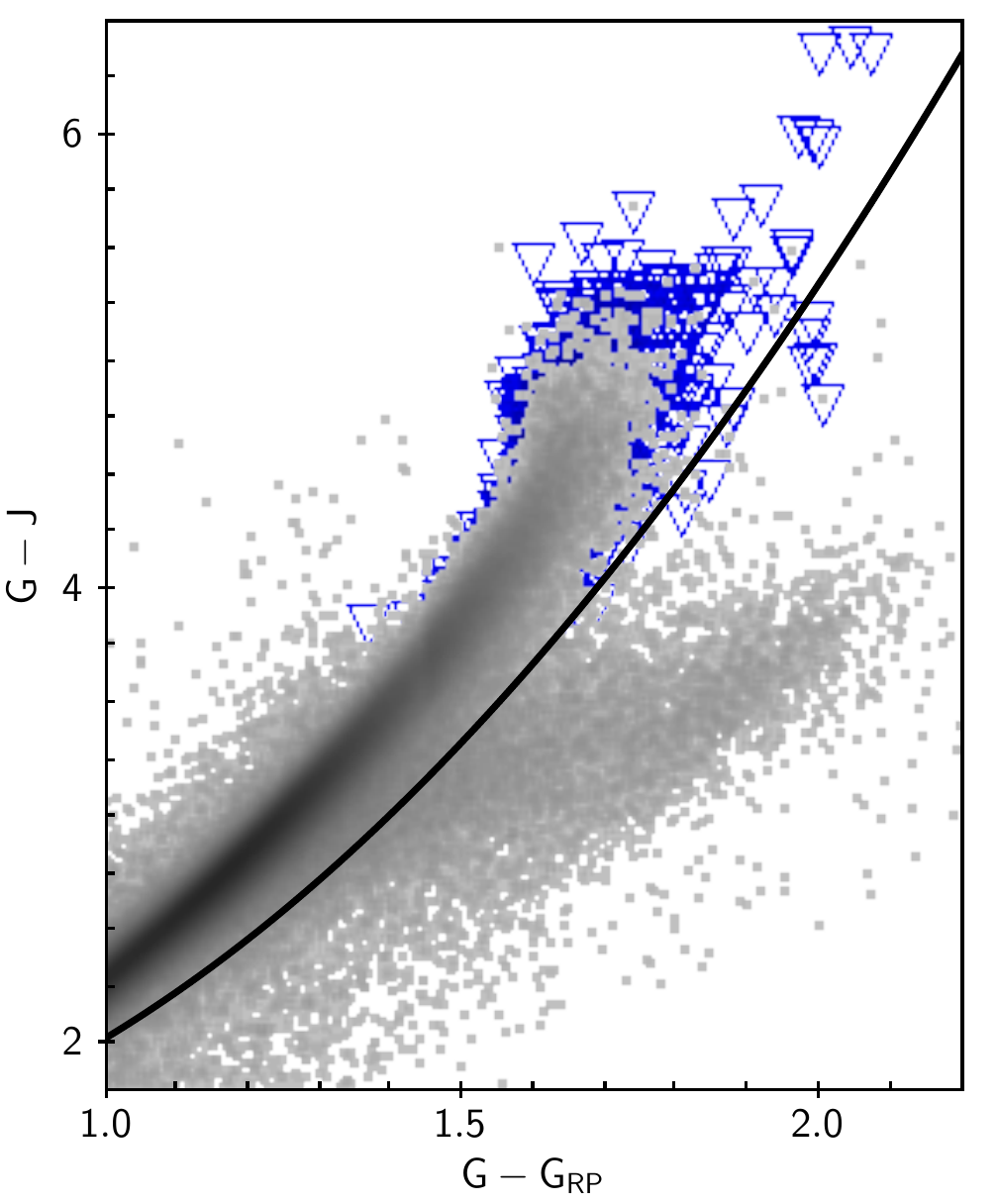}   
\includegraphics[height=7.5cm]{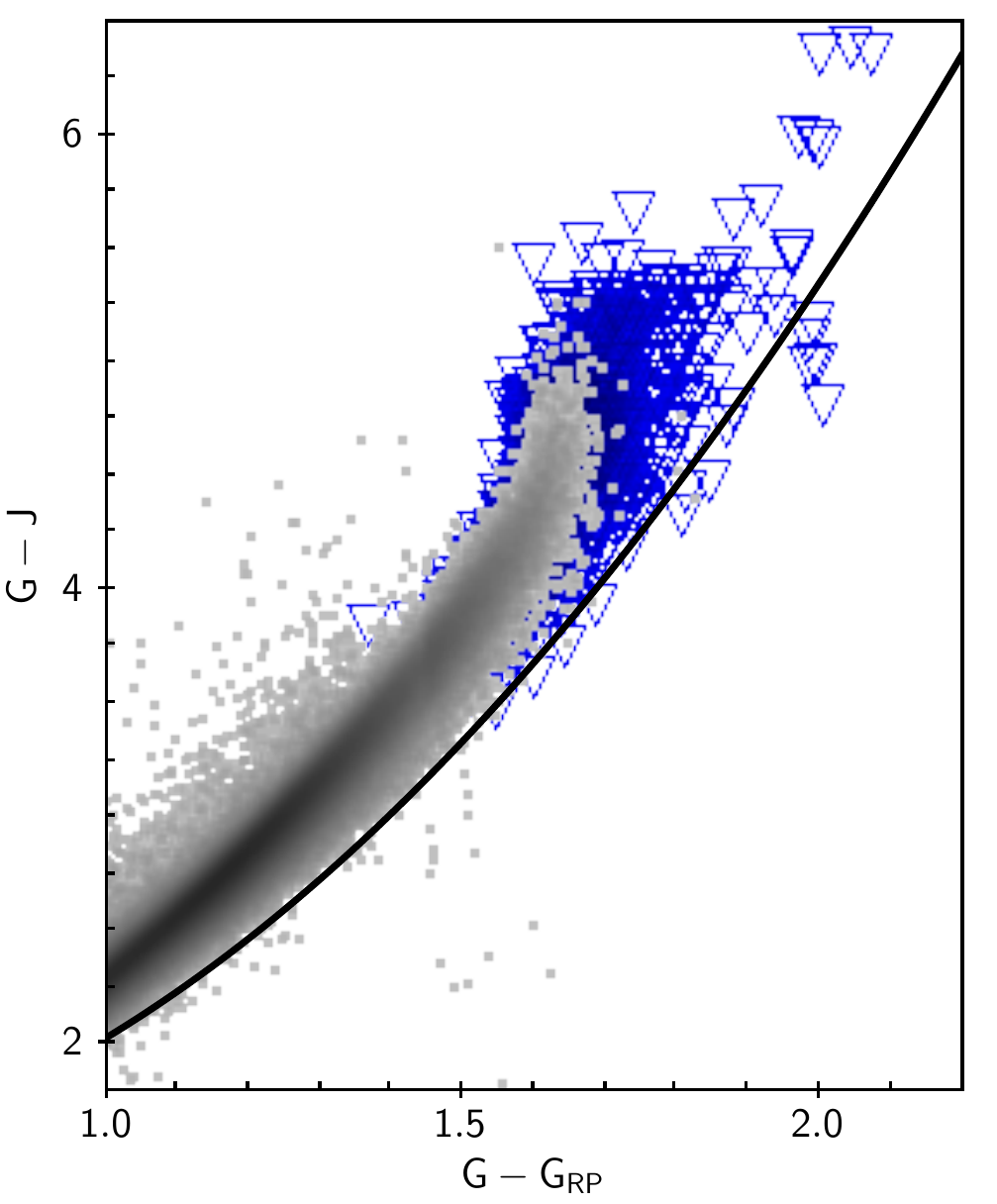}   
\caption{G-J vs. G-\grp diagram at the low-mass end. The open triangles show previously known objects found in \gdrtwo (see  Sect.~\ref{known}). The dots show the filtered \gdrtwo catalog with a 2MASS counterpart and a 2MASS photometric quality flag Q$_\mathrm{fl}$ = AAA, before (left), and after (right) removing objects with an RP/BP flux excess following \cite{Evans2018}. The curve shows the empirical limit we used to filter objects with spurious colors.}
\label{f:colexc}
\end{center}
\end{figure*}

It is thus necessary to define a more suitable criterion to retain these faint and red objects that does not rest upon the BP flux. The effect of color excess is also clearly visible in the G-J versus G-\grp plane (Fig.~\ref{f:colexc}, left panel). We define a new empirical limit, shown by the curve in the plot: $$\mathrm{G}-\mathrm{J} = 1.42 \times (\mathrm{G}-\mathrm{G}_{\mathrm{RP}})^2-0.94\times (\mathrm{G}-\mathrm{G}_{\mathrm{RP}} )+1.55.$$
We note that the J magnitude rejects $\sim$20\% of the candidates with no 2MASS counterpart or a 2MASS photometric quality flag Q$_\mathrm{fl}$ $\neq$ AAA. Removing objects below this limit allows excluding objects with spurious colors, but the low-mass objects can still be retained, as well as most of the previously known sample. This is in contrast to the cut using the color excess factor (Fig.~\ref{f:colexc}, right panel). The resulting HR diagram is shown in the right panel of Fig.~\ref{f:hrddr2}.

As the rejection of objects is based on a hard and empirical cutoff, we expect our sample to be contaminated, in particular for G$-\mathrm{G}_\mathrm{RP} \lesssim 1.6$ (see left panel of Fig.~\ref{f:colexc}). $\text{About}$ 12\% of our sample with G$-\mathrm{G}_\mathrm{RP} \lesssim 1.6$ indeed have  a color-excess factor  above the limit defined by \cite{Evans2018}. This is much less an issue for L candidates (G$-\mathrm{G}_\mathrm{RP} \gtrsim 1.6$).

\subsection{Selecting ultra-cool and brown dwarf candidates}

\begin{figure*}[ht!]
\begin{center}
\includegraphics[height=7.5cm]{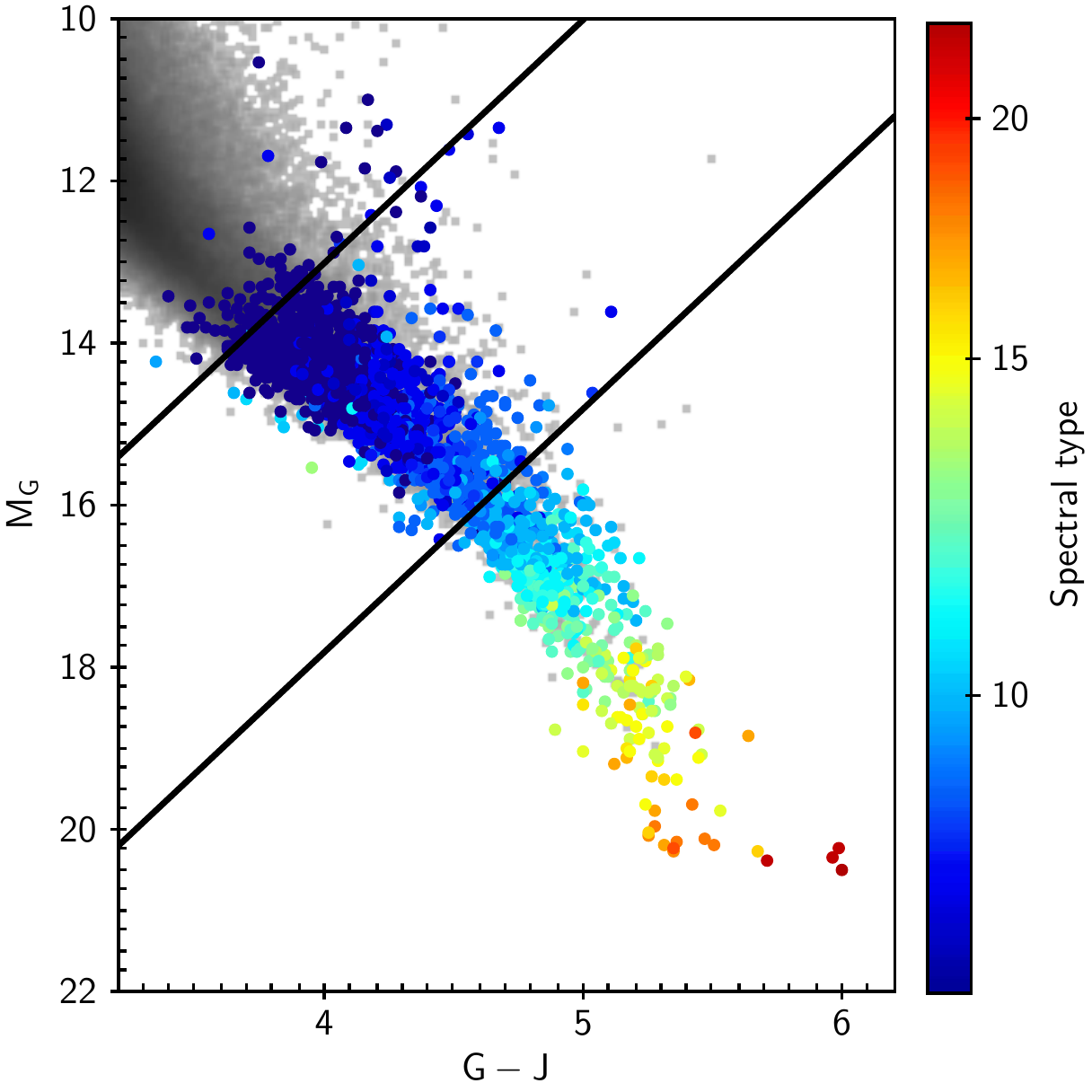}   
\includegraphics[height=7.5cm]{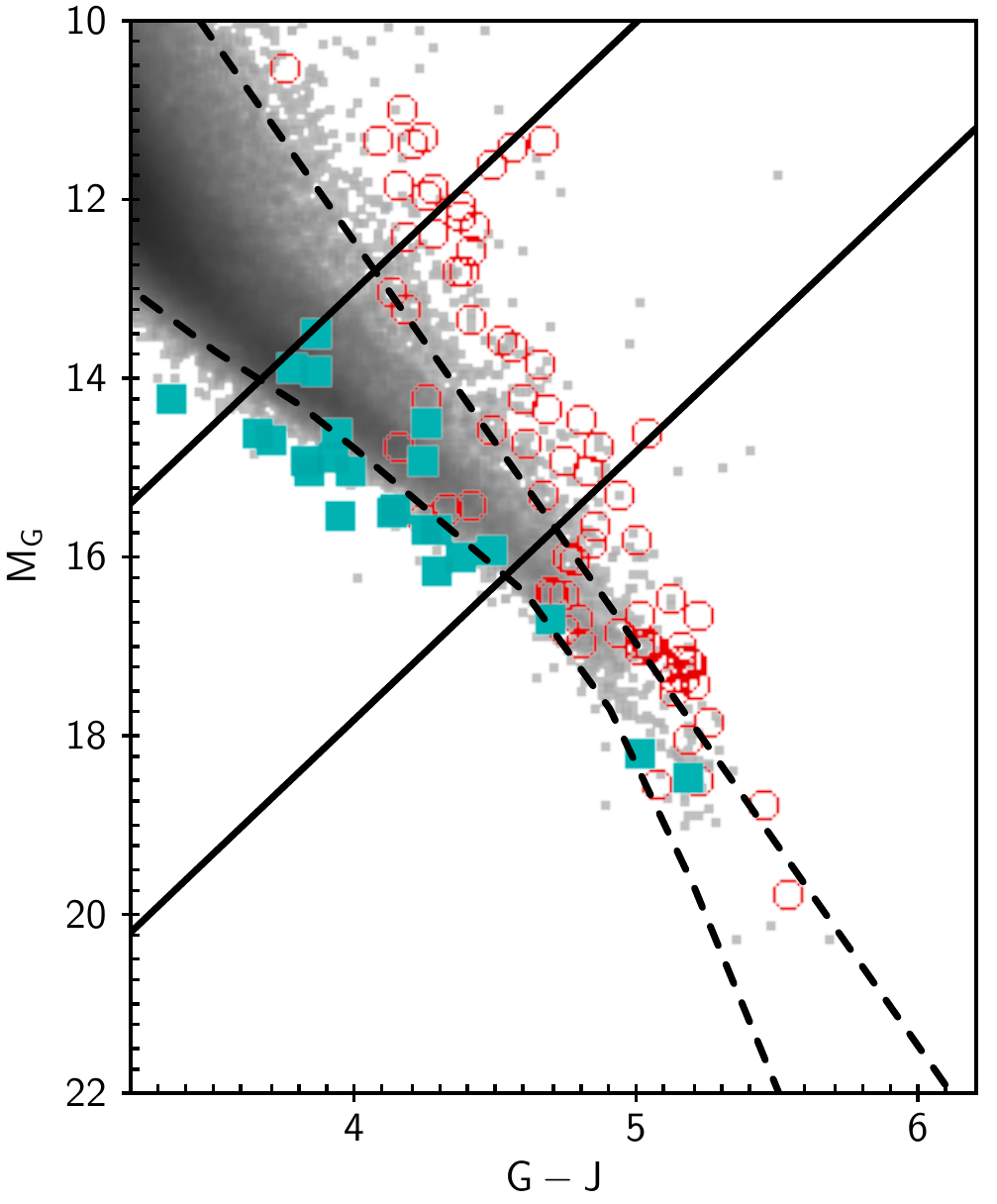}   
\caption{\mg vs. G-J diagram of the filtered \gdrtwo catalog (gray dots) at the low-mass end, superimposed on the known spectroscopic
sample. Left: The color-code gives the spectral type (10 stands for L0 and 20 for T0) and allows defining three
regions, separated by the solid lines, from top to bottom: $\lesssim$M7, $\sim$M7 to L0, and $\gtrsim$L0. Right: Squares represent the known subdwarfs,
and open circles show the young candidates. The dashed lines aim to separate these different types.}.
\label{f:selection}
\end{center}
\end{figure*}

Fig.~\ref{f:selection} focuses on the \gdrtwo\ HR diagram at the low-mass end, superimposed with the sample of spectroscopically confirmed ultra-cool and brown dwarfs found in \gdrtwo. 

We used the information on the spectral type to define separations in the HR diagram. They are shown by solid lines and suggest that \gdrtwo\ contains 14 176  M7 to L0 candidates and 488 L candidates, all of them earlier than L5. 
These indicative numbers exclude previously known objects in common that remained after we applied the filters described in the previous section (870 M7 to L0 and 223 L  dwarfs).

Because the very end on the main sequence is less affected by the color excess problem, we also selected L candidates from the \mg versus G-\grp diagram. The selection was defined from the locus of the spectroscopically confirmed sample  (\mg$>17.5-$(G-\grp), and 1.55 $<$\grp$<$1.85). It allowed us to retrieve robust candidates  that do not depart from the main sequence. These 251 additional candidates would have been excluded based on lack of J magnitude (or a bad 2MASS photometric quality flag Q$_\mathrm{fl}$).

We also used the peculiarities of the known sample to tentatively define regions where subdwarfs and young candidates are expected to be dominant. We used the G-J color index as it better separates these types (right panel in Fig.~\ref{f:selection}, also visible in the right panel of  Fig.~\ref{f:hrdmltvtan}), probably because of the larger wavelength extent between the G and J bands.   There are 233 ($\geq$ M7) and 70 (L) new promising young candidates and 466 ($\geq$ M7) and 17 (L) subdwarf candidates.
The candidates are listed in a table published online\footnote{The full table is only available in electronic form
at the CDS via anonymous ftp to cdsarc.u-strasbg.fr (130.79.128.5) or via http://cdsweb.u-strasbg.fr/cgi-bin/qcat?J/A+A/}. A small part is shown in Table~2.

\begin{table*}[ht]
\label{appendtable}
\caption{Characteristics of the ultra-cool and brown dwarf candidates found in \gdrtwo. 
The full table is available online. 
SpT is the photometric spectral type computed from the \mg vs. spectral type relation defined in Sect.\ref{known}.}

\small
\begin{tabular}{p{2.3cm}llllllllll}
\hline \\
DR2 source ID &  $\alpha$ &      $\delta$ &      parallax &      $\mu_\alpha$ &         $\mu_\delta$ &         G &     G-\grp &       V$_{\mbox{T}}$  &        SpT &   peculiarity \\
& & &mas &mas yr$^{-1}$ &mas yr$^{-1}$ & & &km s$^{-1}$& &\\
\hline \\
741002118375424 &   47.455 &   2.223 &  10.192 &  -52.550 &  -120.279 &  19.648 &   1.554 &  61.046 &   M7.5 &     \\
1577597323028864 &   43.690 &   2.836 &   7.061 &  -42.770 &  -52.972 &  19.393 &   1.482 &  45.702 &   M7.0 &     \\
2038460198792704 &   44.559 &   3.795 &  10.712 &  33.802 &  27.336 &  19.009 &   1.468 &  19.237 &   M7.0 &     \\
2781562554898432 &   47.779 &   4.293 &  23.509 &  399.310 &  -34.376 &  18.561 &   1.596 &  80.810 &   M8.0 &     \\
3036954195695616 &   49.836 &   4.733 &  18.310 &  75.223 &  -75.181 &  18.453 &   1.559 &  27.532 &   M7.5 &     \\
3447449989685760 &   46.704 &   4.426 &   9.889 &  -12.907 &  -52.790 &  20.289 &   1.578 &  26.050 &   M8.0 &     \\
5797037618821888 &   41.487 &   4.533 &  11.794 &  90.101 &  -96.024 &  19.145 &   1.588 &  52.921 &   M7.5 &     \\
6033570057616896 &   40.247 &   5.165 &   7.469 &  38.655 &  -28.239 &  19.817 &   1.675 &  30.381 &   M7.0 &     \\
6078688688994048 &   40.837 &   5.131 &  15.053 &  337.776 &  -36.942 &  18.198 &   1.485 &  106.995 &   M7.0 &     \\
6156543561459584 &   40.374 &   5.613 &   9.231 &  40.170 &   3.470 &  18.874 &   1.497 &  20.704 &   M7.0 &     \\
6313262622741504 &   42.279 &   5.720 &  11.231 &  158.451 &  -51.802 &  18.429 &   1.477 &  70.357 &   M7.0 &     \\
6668160065596928 &   42.433 &   6.508 &  10.274 &  -63.768 &  -59.994 &  18.890 &   1.483 &  40.394 &   M7.0 &     \\
7033094847307136 &   45.649 &   6.409 &   8.880 &  37.677 &  -43.789 &  19.593 &   1.471 &  30.834 &   M7.0 &  subdwarf \\
8052342126139392 &   44.378 &   7.053 &  13.937 &  66.670 &  -62.054 &  20.122 &   1.648 &  30.976 &   M9.0 &     \\
8777268181161216 &   44.872 &   8.471 &  25.662 &  158.484 &  13.082 &  16.794 &   1.452 &  29.373 &   M7.0 &     \\
9475659927993600 &   51.236 &   6.520 &   9.305 &  50.899 &  -27.089 &  18.865 &   1.509 &  29.371 &   M7.0 &     \\
9790158907981952 &   52.557 &   7.060 &   8.240 &  73.355 &  -14.970 &  19.373 &   1.460 &  43.068 &   M7.0 &     \\
9851005710350592 &   52.683 &   7.727 &   9.143 &  21.240 &  -33.522 &  19.173 &   1.490 &  20.573 &   M7.0 &     \\
10358086728974848 &   49.722 &   7.165 &  13.882 &  53.955 &  -52.929 &  18.110 &   1.457 &  25.808 &   M7.0 &     \\
12880641280670976 &   51.821 &  10.057 &  12.777 &  55.051 &  -38.696 &  19.631 &   1.557 &  24.963 &   M8.0 &   young  \\
15647768450759296 &   47.629 &  11.203 &  22.431 &  110.190 &  -371.818 &  17.788 &   1.522 &  81.947 &   M7.5 &     \\
16744020197785216 &   51.278 &  12.472 &  10.185 &  74.968 &  -7.001 &  19.201 &   1.573 &  35.042 &   M7.0 &     \\
17275870293313024 &   48.488 &  12.065 &  11.908 &  63.823 &  -187.191 &  18.668 &   1.492 &  78.721 &   M7.0 &     \\
17315212193436672 &   49.131 &  12.133 &  13.993 &  56.529 &   6.926 &  18.244 &   1.494 &  19.292 &   M7.0 &     \\
18771343546171008 &   41.647 &   7.080 &  11.915 &  88.864 &  -14.731 &  18.611 &   1.495 &  35.833 &   M7.0 &     \\
19314472225770368 &   38.677 &   7.087 &  10.064 &  -82.286 &  -68.344 &  18.734 &   1.470 &  50.379 &   M7.0 &     \\
19350511296260224 &   38.990 &   7.412 &   6.656 &  -34.833 &  -20.406 &  19.357 &   1.463 &  28.747 &   M7.0 &     \\
19485927320169472 &   37.682 &   7.298 &  10.811 &  20.122 &  -14.774 &  18.669 &   1.549 &  10.945 &   M7.0 &     \\
19499701279778432 &   36.956 &   7.118 &  17.641 &  94.693 &  164.794 &  18.328 &   1.551 &  51.069 &   M7.5 &     \\
19697033552145920 &   38.025 &   8.207 &  23.744 &  171.769 &  11.269 &  17.729 &   1.524 &  34.364 &   M7.5 &     \\
\hline \\
\end{tabular}
\end{table*}

\subsection{Comparison with models}

Evolutionary and atmosphere models of ultra-cool and brown dwarfs have been developed for a long time \citep[see, e.g., the 20-year old review from][]{Burrows1998}
and are successful today in reproducing color-magnitude diagrams in many photometric systems.

However, the strong added value of \gaia\ data in testing these models, that is, the precise parallax, will help refine these models. Here we show a preliminary comparison of our candidates with the new evolutionary models reported by \cite{Baraffe2015} that consistently couple interior structure calculations with the BT-Settl atmosphere models \citep{Allard2013}.

\begin{figure*}[ht!]
\begin{center}
\includegraphics[height=7.5cm, clip=, viewport=2 0 290 350]{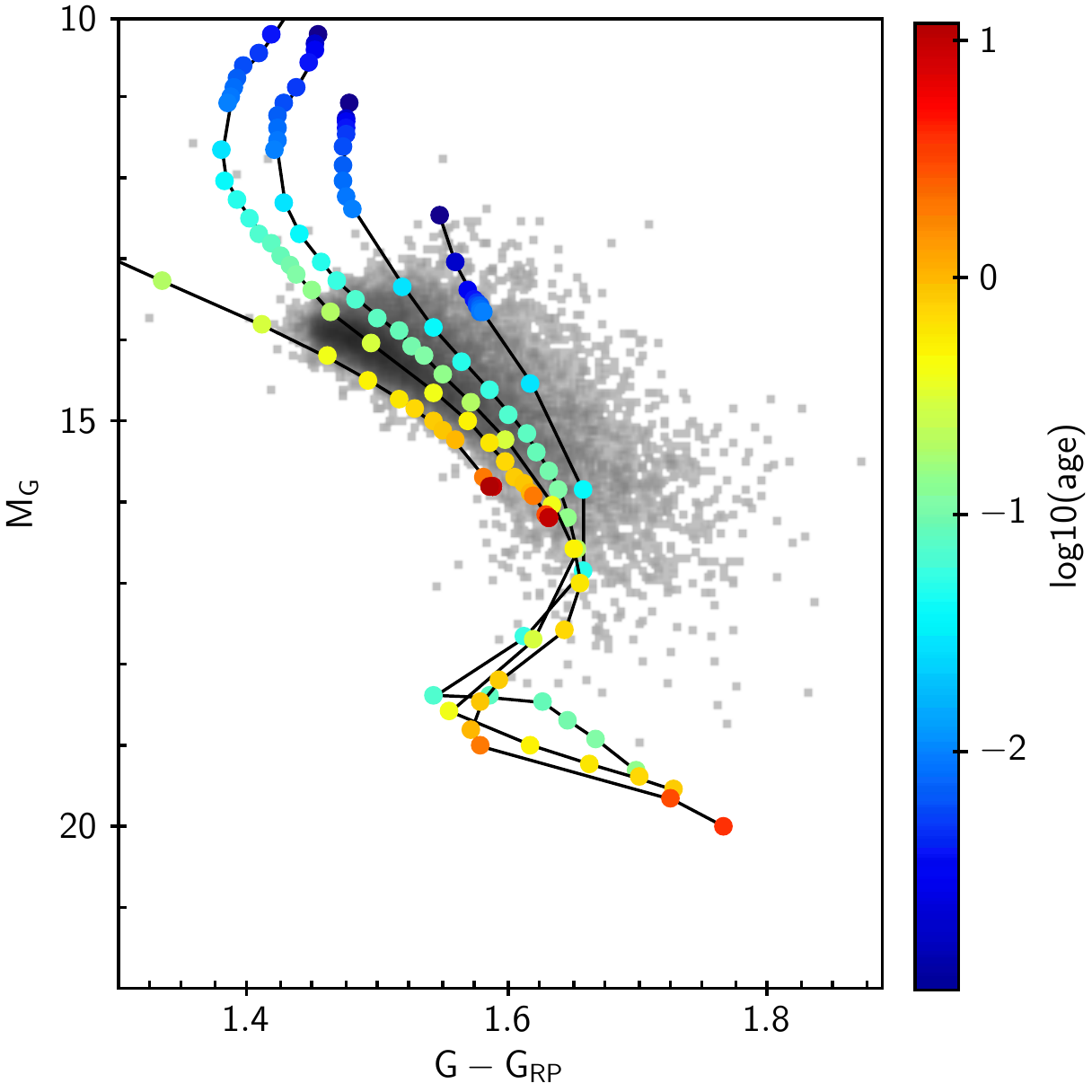}   
\includegraphics[height=7.5cm]{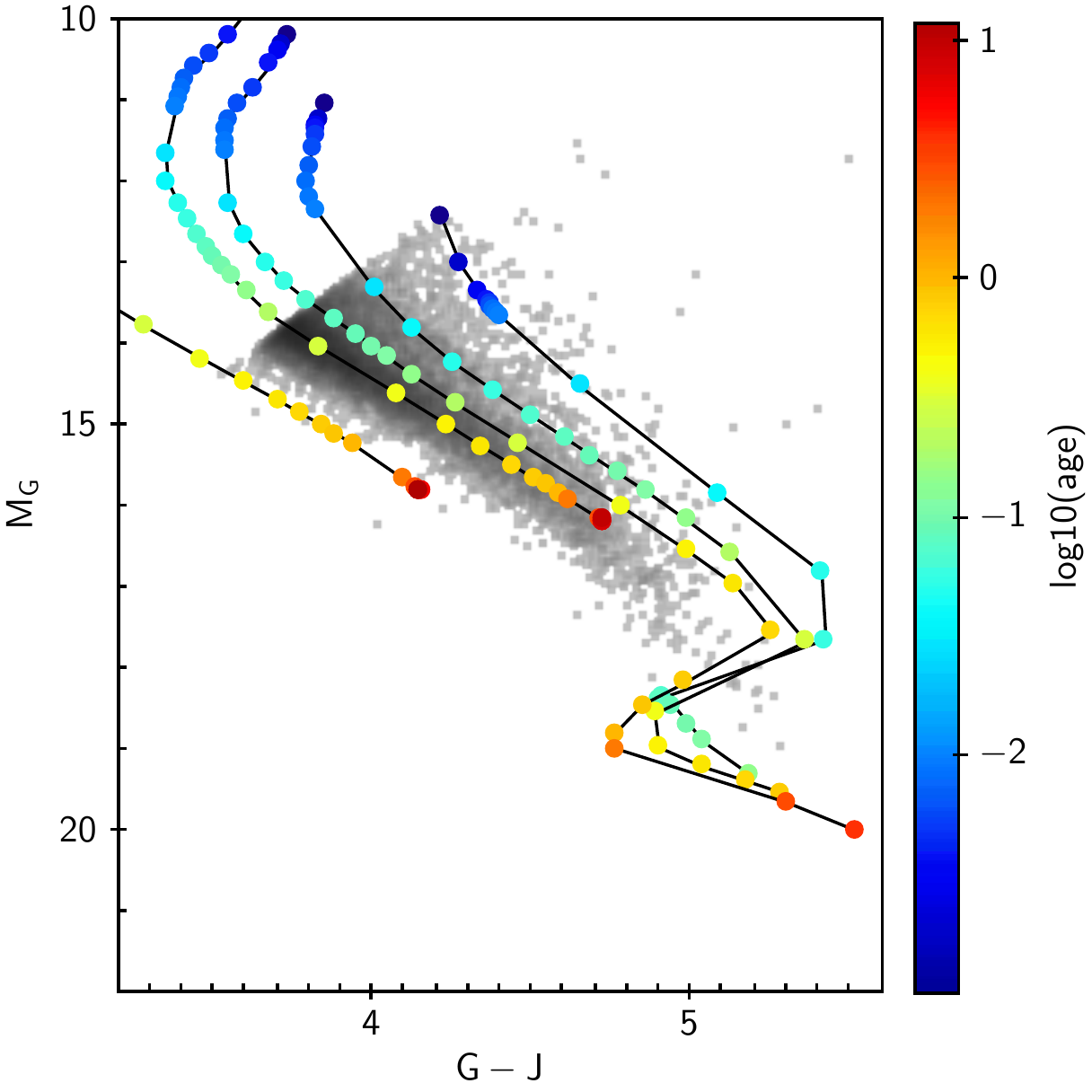}   
\caption{BT-Settl evolution tracks
from right to left: 0.02, 0.04, 0.06, and 0.08 M$_\odot$ at [M/H]=0, and 0.083 M$_\odot$ at [M/H]=-1. The color bar gives the logarithm of the age in Gyr. The gray dots show the \gdrtwo candidates.}
\label{f:model}
\end{center}
\end{figure*}

Fig.~\ref{f:model} shows the HR diagram of our candidates superimposed on these evolutionary models. They tend to confirm the locus of the objects as a function of the peculiarities we discussed in the previous section.  It also shows that our selection does not favor young candidates, as can  be seen also in the right panel of Fig.~\ref{f:selection}.

\subsection{Completing the census}

\begin{figure}[ht!]
\begin{center}
\includegraphics[height=5cm]{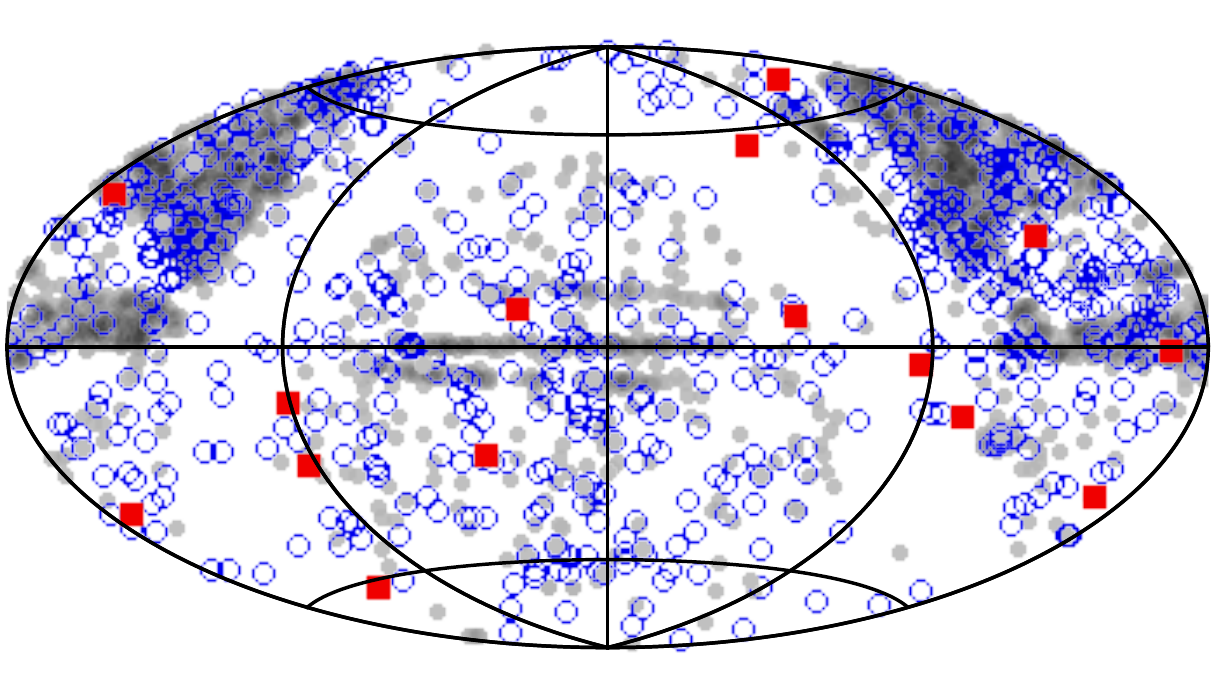}   
\includegraphics[height=5cm]{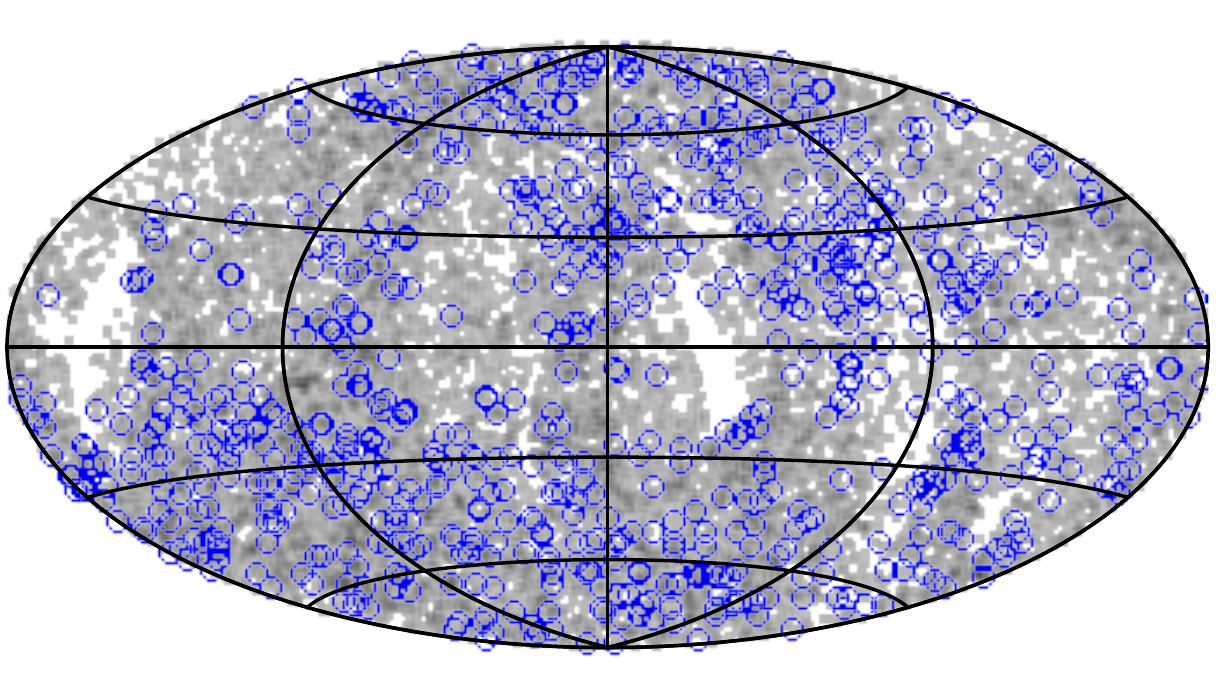}      
\caption{Sky distribution in equatorial coordinates of previously known ultra-cool dwarfs with a counterpart in \gdrtwo\ (top), and of the additional candidates found in  the filtered \gdrtwo\ data (bottom). The dots are $\geq$ M7, the blue open circles are L-dwarfs, and the red squares are T-dwarfs.}
\label{f:aitoff}
\end{center}
\end{figure}

Fig.~\ref{f:aitoff} shows the sky distribution of previously known (top) and new candidates (bottom). The new candidates cover the whole sky, including the Galactic plane, where we can expect more discoveries of hidden dwarfs, such as the newly characterized L-dwarf WISE J192512.78+070038.8 \citep{Scholz2018,Faherty2018}. The imprint of the \gaia scanning law is visible in the plot, where no candidates are found in some Galactic regions, which are rejected as they do not fulfill our condition \texttt{visibility$\_$periods$\_$used} $>8$. Thus new candidates are expected to be found in the forthcoming \gaia data releases.

\begin{figure*}[ht!]
\begin{center}
\includegraphics[height=7.5cm]{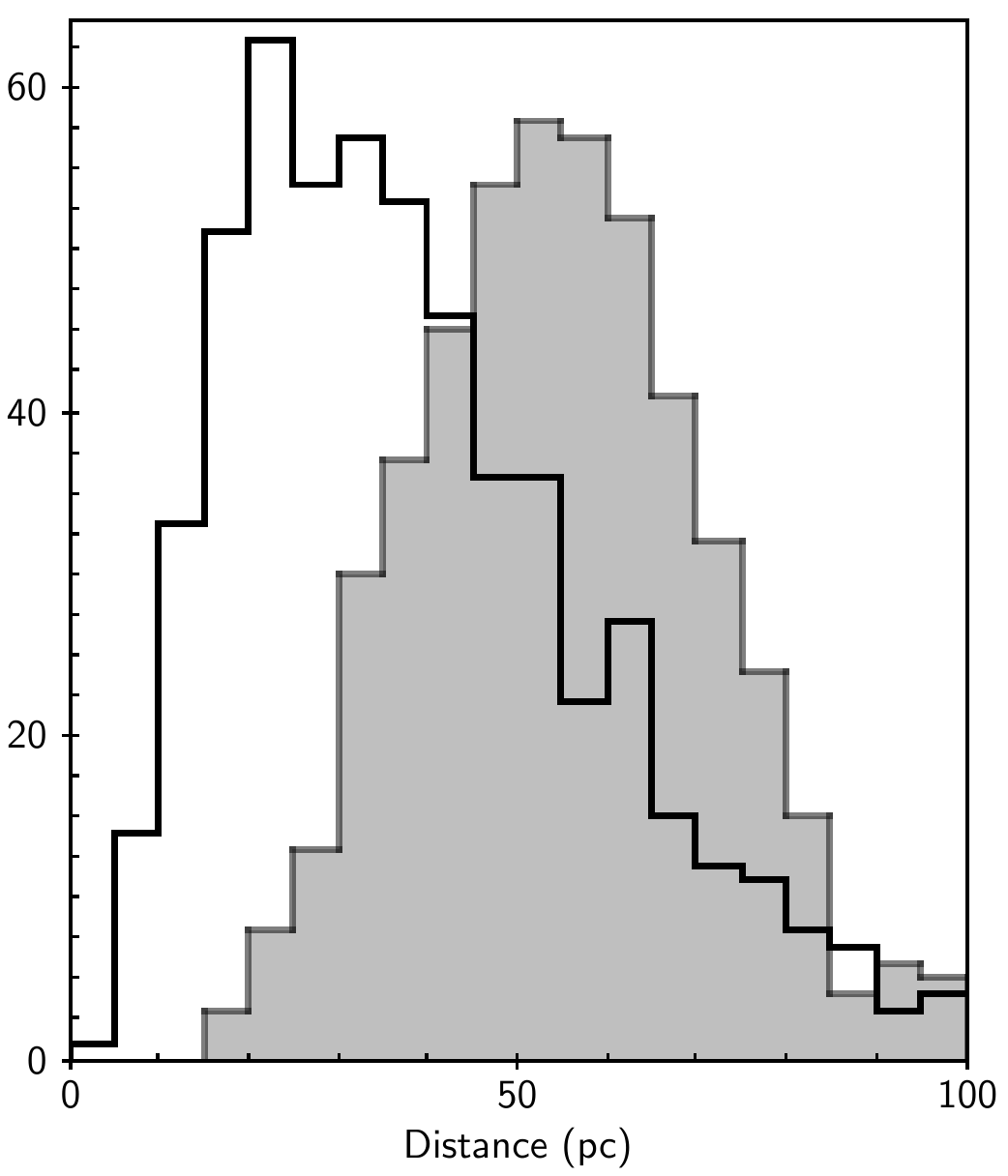}   
\includegraphics[height=7.5cm]{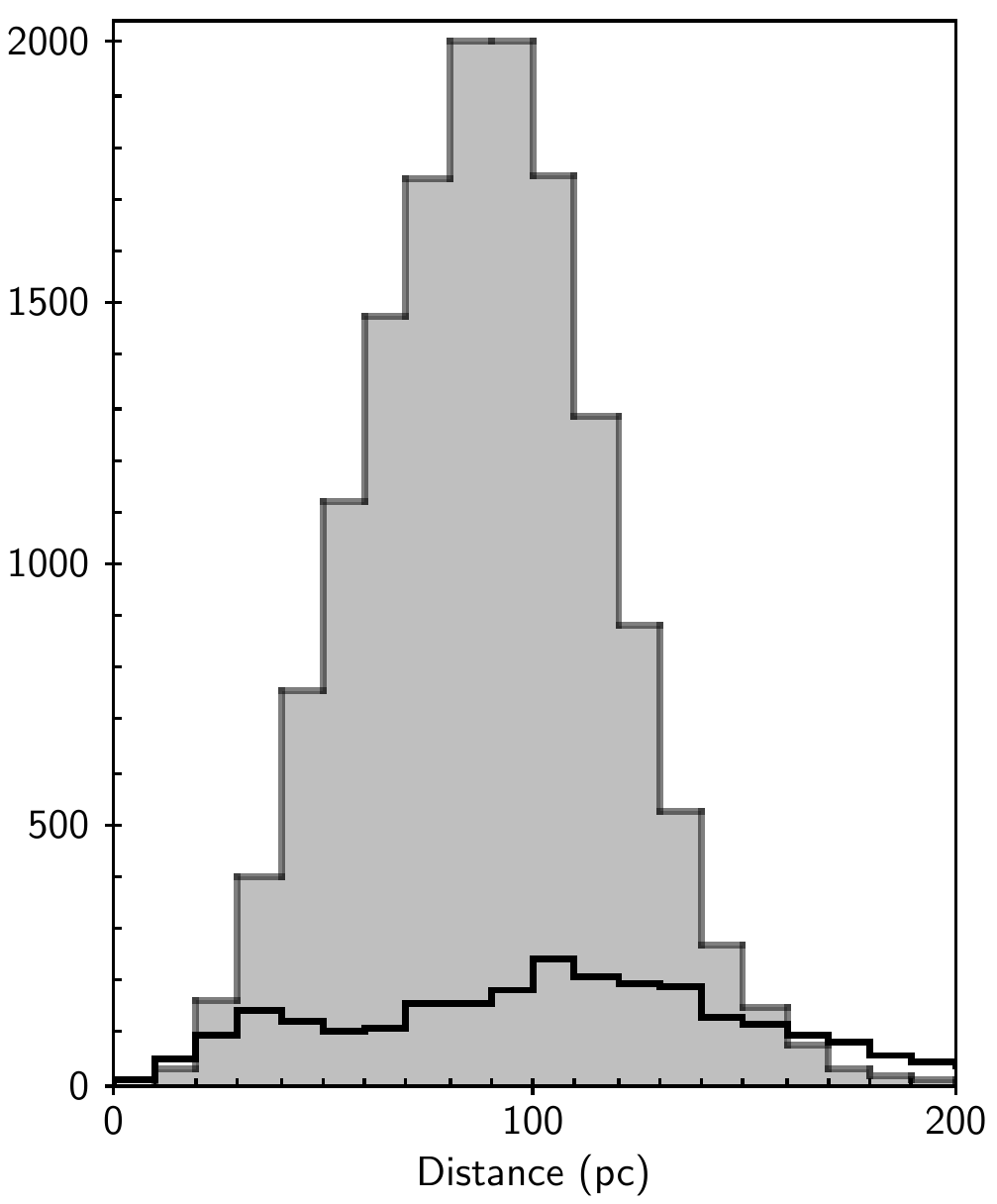}   
\caption{Distance distribution of the L (left) and $\geq$M7 to L0 (right) dwarfs for the known sample (black line) and the additional candidates found in the filtered \gdrtwo\ data (filled gray).}
\label{f:histdist}
\end{center}
\end{figure*}

Fig~\ref{f:histdist} shows the distance distribution of previously known and additional \gdrtwo candidates  for L and $\geq$ M7 dwarfs.
While several recent efforts \citep[see][]{Best_aas2018} have been made to complete the volume-limited sample of brown dwarfs up to 25 pc, \gaia\ shows that the census is still not complete. The situation is even worse for $\geq$M7 dwarfs, which is not surprising since we here only considered objects with spectroscopic confirmation. The ongoing and future spectroscopic large survey such as APOGEE \citep{Majewski2017}, LAMOST \citep{Cui2012}, WEAVE \citep{Dalton2014}, and 4MOST \citep{Dejong2016} will soon complete the characterization of the census of sources that lie farther away.

\section{Conclusion}
\label{ccl}

\cite{Smart2017} estimated the \gaia magnitude from J magnitude and spectral type of spectroscopically confirmed objects and found that 1010 L and 58 T  are brighter than G$_\mathrm{est} = 21.5$,  and 543 L and 10 T  are brighter than G$_\mathrm{est} = 20.3$.  \cite{Sarro2013} predicted similar numbers using a more theoretical approach.

\gdrtwo\ contains very many objects of the spectroscopically confirmed sample: 3050 ultra-cool ($\geq$M7), 647 L, and 16 T dwarfs. This corresponds to 61\% and 74\% objects  with G$_\mathrm{est} \geq 21.5$ and G$_\mathrm{est} =\geq20.3$, respectively. However, \gdrtwo\ is an intermediate release, on which strong filters were applied. Further releases will most probably continue to fill the gap.

There are numerous ultra-cool ($\sim$14 200) and mainly early L (739) candidates in \gdrtwo. We stress that these numbers are lower limits because of the successive cuts we made to the original \gdrtwo. In particular, we used the J magnitude to remove objects with a spurious color excess, and  $\sim$ 20\% of the objects were rejected during the cross-identification with 2MASS. The high precision of the HR diagram allows us to give an indication of some of the characteristics of the object, for instance, whether it is young or a subdwarf. This rough classification was made to guide the target selection for follow-up, depending on the scientific case that is to be studied. 

The given numbers are of course indicative, but they also prove that the census is not complete yet, even locally.
Although we call them new candidates, some may be part of the significant number of ultra-cool and brown dwarf candidates found in large-scale surveys that have been classified from photometry \citep[e.g., ][]{Folkes2012,Smith2014,Smith2018,Schmidt2015,Skrzypek2016,Theissen2017,Best2018}. We cross-matched these studies and found that 1800 $\geq$ M7 and 108 L dwarfs are indeed part of them. Still, the numerous candidates we found are most valuable for follow-up since they benefit from the reliable \gaia\ observations, including precise parallax and proper motion.

Further exploitation of this dataset will be useful for completing the census of late-M and at least early-type brown dwarfs, refining their luminosity function, testing stellar and substellar models, and probing the old population in the Galaxy through the few hundred subdwarf candidates. The contamination and completeness of the sample is not easy to address, in particular because of the BP and RP flux measurements, which are contaminates by the background flux. The cyclic nature of the treatment of \gaia offers the promise of a increased quality and a larger quantity of data at each  release.
In any case, the high precision of \gaia\ and the numerous candidates it provides with full 5D (astrometry and photometry) data offers a great wealth of information at the low-mass end of the HR diagram. This homogeneous and precise catalog is a worthwhile target to be further characterized with spectroscopic observations, and most likely has surprises in store for us.

\begin{acknowledgements}
The author thanks the referee for fruitful comments. This work has made use of data from the European Space Agency (ESA) mission
{\it Gaia} (\url{https://www.cosmos.esa.int/gaia}), processed by the {\it Gaia}
Data Processing and Analysis Consortium (DPAC,
\url{https://www.cosmos.esa.int/web/gaia/dpac/consortium}). Funding for the DPAC
has been provided by national institutions, in particular the institutions
participating in the {\it Gaia} Multilateral Agreement. This research has made use of the VizieR catalogue access tool, CDS, Strasbourg, France. The original description of the VizieR service was published in A\&AS 143, 23. This research has made use of the SIMBAD database,
operated at CDS, Strasbourg, France. The author made  queries at CDS using Virtual Observatory with python \citep{Paletou2014}. All figures have been generated using the TOPCAT\footnote{http://www.starlink.ac.uk/topcat/} tool \citep{Taylor2005}. Data fitting was performed using Gnuplot\footnote{http://www.gnuplot.info/}.
\end{acknowledgements}

\bibliographystyle{aa}
\bibliography{BDgaia.bib}

\end{document}